# In-Plane Electric Field Induced Orbital Hybridization of Excitonic States In Monolayer $WSe_2$

Bairen Zhu [1,2,+,*], Ke Xiao [2,+], Siyuan Yang [2], Kenji Watanabe [3], Takashi Taniguchi [4], Xiaodong Cui [2,*]

[1]. Department of Applied Physics, Zhejiang University of Technology, Hangzhou 310023, China

[2]. Physics Department, University of Hong Kong, Hong Kong, China

[3]. Research Center for Functional Materials, National Institute for Materials Science, 1-1 Namiki, Tsukuba 305-0044, Japan

[4]. International Center for Materials Nanoarchitectonics, National Institute for Materials Science, 1-1 Namiki, Tsukuba 305-0044, Japan

[+] The authors contribute equally.

[*] Email: zhubair@zjut.edu.cn and xdcui@hku.hk

**Abstract**

The giant exciton binding energy and the richness of degrees of freedom make monolayer transition metal dichalcogenide an unprecedented playground for exploring exciton physics in 2D systems. Thanks to the well energetically separated excitonic states, the response of the discrete excitonic states to the electric field could be precisely examined. Here we utilize the photocurrent spectroscopy to probe excitonic states under a static in-plane electric field. We demonstrate that the in-plane electric field leads to a significant orbital hybridization of Rydberg excitonic states with different angular momentum (especially orbital hybridization of 2$s$ and 2$p$) and consequently optically actives 2$p$-state exciton. Besides, the electric-field controlled mixing of the high lying exciton state and continuum band enhances the oscillator strength of the discrete excited exciton states. This electric field modulation of the excitonic states in monolayer TMDs provides a paradigm of the manipulation of 2D excitons for potential applications of the electro-optical modulation in 2D semiconductors.

Hybridization primitively indicates a mixing of atomic orbitals of comparable energies for the description of chemical bonds in valence band theory. In 2D TMDs, band edges around ±K valleys are primarily constructed by *d* orbitals of transition-metal atoms with a small component of *p* orbitals of chalcogen atoms due to atomic orbital hybridization[1-3]. This atomic orbital hybridization leads to a significant spin-orbit coupling in the conduction band edge[2]. Orbital hybridization of valley excitons exists not only among atomic orbitals but also in excitonic orbitals. For valley excitons in monolayer TMDs, the complete wavefunction is described with the direct product of electron/hole's Bloch functions featured with parent atomic orbitals and envelope functions featured with excitonic orbitals. The envelop functions denote the relative motions between electron and hole and could be described with a Rydberg notation ($n,l$), where *n,l* are the principal index ($n$=1,2,3…) and orbital angular momentum quantum number ($l$=0,1,2···), respectively. The different Rydberg excitonic orbitals have the distinct optical properties, in an over-simplified manner, either bright or dark. The bright excitons denote the optically active excitons associated with the dipole allowed one-photon transition which dominates the linear optical properties. They are usually with a *s*-state exciton envelop function with zero orbital angular momentum ($l$=0). However, their excited states are not necessarily bright. For example, 2*p*, 3*p*,…-state excitons are naturally considered as the angular-momentum forbidden dark excitons owing to the symmetry of the envelop function in the standard 2D exciton model[4-7]. In monolayer TMDs, the three-fold in-plane rotational symmetry from the crystal lattice renders *p*-state excitons bright with the microscopic mechanism of the trigonal warping effect[8]. However, the *p*-state excitons are absent in the experimental linear optical spectra[9-11] and the calculations[8,12] attribute the dark *p*-state excitons to the negligible oscillator strength in one-photon (dipole allowed) transition. Recently, there are several theoretical proposals that disorders, Rashaba spin-orbit interaction or skyrmions could brighten *p*-state excitons[13,14].

Here we demonstrate an electric-field controlled excitonic orbital hybridization in *h*BN-encapsulated monolayer $WSe_2$ with the linear photocurrent spectroscopy. We observe that an in-plane electric field could break the rotational symmetry of exciton envelop functions and induce a significant orbital hybridization of Rydberg excitons with different angular momentum by mixing their envelope functions and consequently brighten 2*p*-state exciton. Besides, the hybridization between high lying excitons and continuum band enhances the oscillator strength of the discrete exciton states. Our experiments unambiguously demonstrate that the energy, linewidth and the oscillator strength of the hybrid exciton states could be efficiently modulated with an in-plane electric field owing to the unique character of 2D excitons.

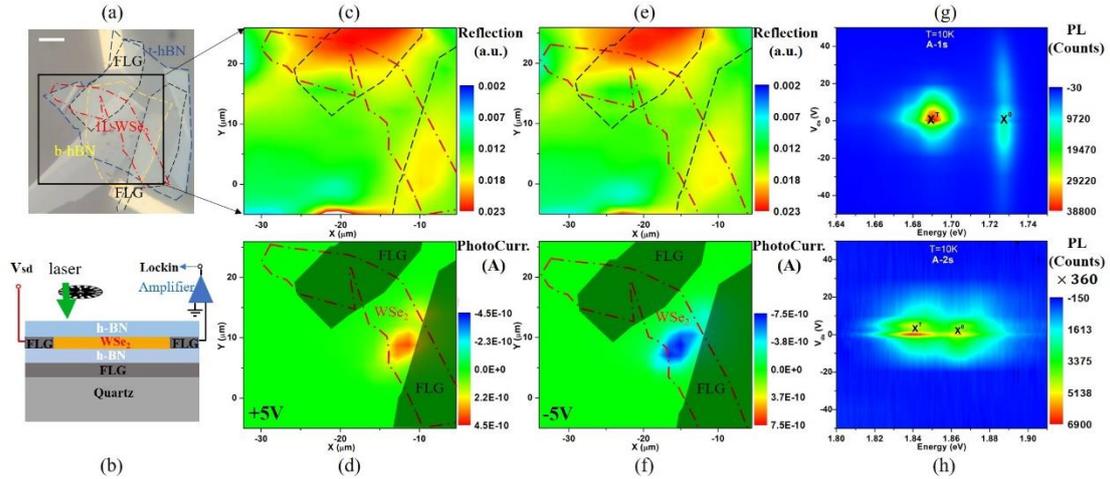

Figure-1 (a) optical microscopic image (Leica DM4 with Sony a6400) of the $h$BN-encapsulated monolayer WSe$_2$ with few-layer graphene (FLG) as electric contacts. The different layers are marked by dash lines of different colors. The white scale bar is 5$\mu m$. (b) Schematic of the device on a transparent quartz substrate for the photocurrent measurement. (c)-(d) the reflection and photocurrent mappings of the device at $V_{ds}$=+5V & T=10K. The scanning range covers the marked area in figure (a). Two electric contacts of FLG and 1L-WSe$_2$ are also labelled for reference. (e)-(f) the reflection and photocurrent mappings at $V_{ds}$=-5V. Photoluminescence of A-1$s$ exciton (g) & A-2$s$ exciton (h) on the spot where the photocurrent is taken as a function of the source-drain voltage at T=10K. The fake colors in these mappings present the signal magnitude with the linear spectra scales where the minimum starts from blue to red at maximum.

The device is made of $h$BN-encapsulated monolayer WSe$_2$ with few-layer graphene as electric contacts. To avoid the photo gating effect from substrates, we transfer the device onto transparent fused quartz substrates, as shown in Fig. 1a. The excitation beam is focused through a scanning objective onto a spot of ~1 micron on the device. Excitons are dissociated by the local electric field and a photocurrent is collected via a lock-in technique (Fig. 1b). Figure 1c-d show the representative reflection and photocurrent maps under $V_{ds}$=+5V excited by a 515nm laser source at T=10K. The photocurrent arises at the region adjacent to the electric contact where the local electric field is expected to be maximum and strong enough to ionize excitons.

The local carrier density is investigated by the 2D color map of low-temperature photoluminescence (PL) as a function of the source-drain bias from -50V to +50V. Figure 1g shows that the PL map of the ground state band edge exciton A-1$s$ (which denotes the ground 1$s$ state of the band edge A exciton) and its corresponding trions peak around $V_{ds}$=0V. The trions consisting of two electron-charged trions, namely intervalley and intravalley trions, signal the weak electron doping at zero bias. The PL peaks redshift away from the zero-bias state, and fade with the increasing source-drain bias in both negative and positive directions. These phenomena are significantly different from those in the electric gated PL experiments, where the increased carrier density transfers the oscillator strength from neutral excitons to trions and shifts the energy of the neutral excitons and trions in opposite directions, as described with the

Fermi-polaron model[15,16]. This in-plane modulated PL experiment reveals the in-plane electric field in force without noticeable doping effect. The suppressed PL emission at the increased bias implies the exciton dissociation under external electric field. The A-1$s$ trions fade at lower electric field than that of A-1$s$ neutral exciton and it is attributed to the small binding energies of the trions. The red-shift of both neutral excitons and trions with the increased electric field could be regarded as the excitonic DC Stark effect[17,18]. The similar electric field dependent PL behaviors occur in A-2$s$ excitons (Fig. 1h) with a higher sensitivity to the in-plane electric field. The PL intensities of both neutral A-2$s$ excitons and the related trions drop to an undetectable level under the bias beyond $V_{ds}=\pm 10V$, as a result of weaker exciton binding energies. Obviously, the PL spectroscopy of Rydberg excitons is not capable to trace the effect of in-plane electric field.

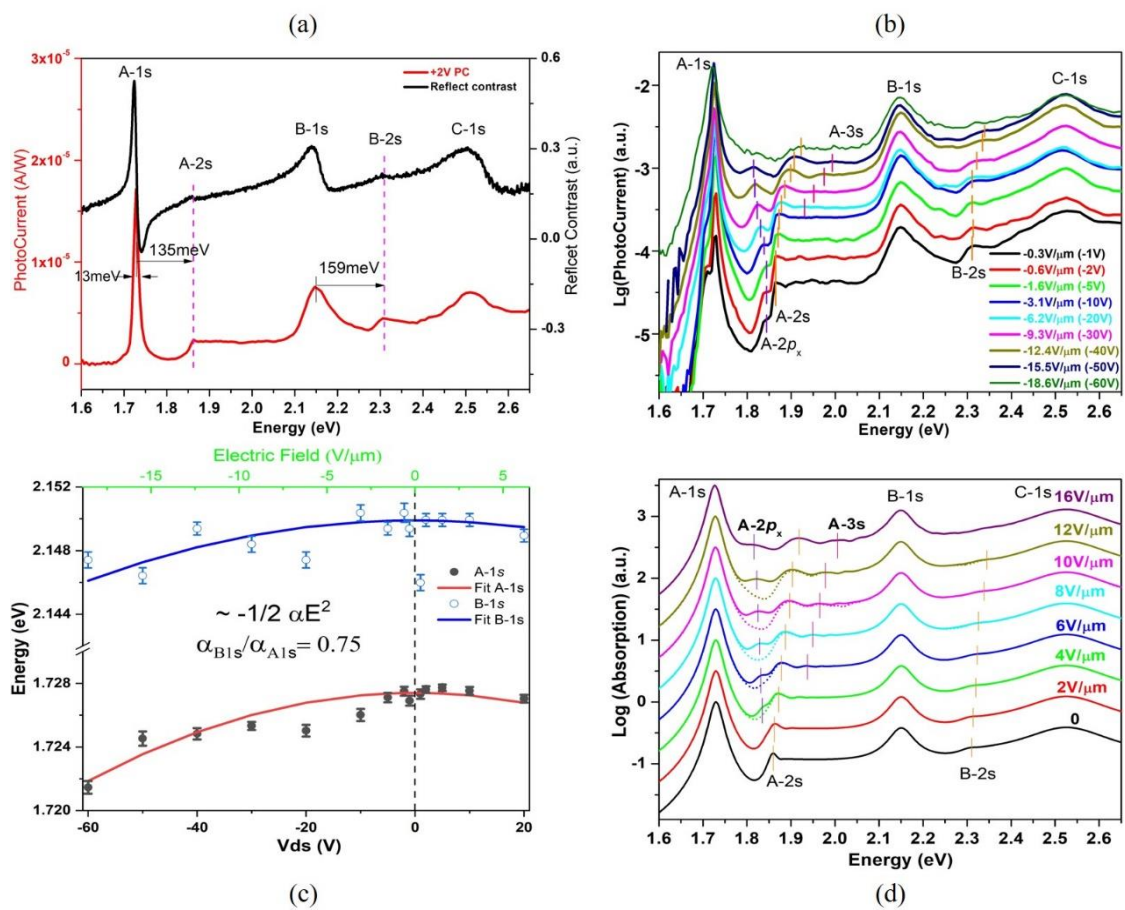

Figure-2 (a) Reflectance contrast and photocurrent spectra at $V_{ds}=+2V$ of $h$BN-encapsulated monolayer WSe$_2$ devices. (b) Photocurrent spectra in log scale under various in-plane electric fields (a zoom-in of the shifts of A1$s$ and the data under the positive electric field are shown in Supplementary Note 2), which are extracted by excitonic DC stark effect of A-1$s$ exciton, as showed in the upper green frame of (c). The energy shift of 1$s$ exciton is proportional to the square of electric field $F$ and exciton in-plane polarizability $\alpha$ ($\sim -1/2\alpha F^2$) with the fitting error bar. The in-plane polarizability of B exciton is found to be smaller than that of A exciton due to a larger exciton

binding energy resulted from the larger effective mass. Similar data from other devices are shown in Supplementary Note 3. (d) The calculated electro-absorption of 1L WSe$_2$ in log scale from 0 to 16V/$\mu m$, consistent with the experimental results in (b). The dash lines denote the electro-absorption spectra without 2*p* state.

We record the reflectance contrast and photocurrent spectra of *h*BN-encapsulated monolayer WSe$_2$ by the excitation energy/wavelength with an intensity below 0.5μW. Figure 2a depicts the reflectance contrast and the photocurrent at V$_{ds}$=+2V as a function of the excitation energy. The band edge ground state exciton A-1*s*, the ground state spin-off exciton B-1*s* and the excitons around the Γ point of the Brillouin zone C-1*s* exciton are clearly observed in both spectra. A significant feature of the photocurrent spectra is the enhanced weight of the excited-state excitons including A-2*s* and B-2*s*. Although the occupation of these excited state is much less than the ground states, the weaker binding energy leads to much higher ionization rate as the electron-hole tunneling is the dominating mechanism for exciton ionization in monolayer TMDs[19,20] (see Supplementary Note 1). So, it is advantageous to study the excited exciton states with photocurrent spectroscopy.

Figure 2a shows that the excited-state excitons A-2*s* and B-2*s* lie at 135, 159meV above their corresponding ground-state excitons A-1*s* and B-1*s*, respectively. It hints that the binding energy of B-excitons is larger than that of A-exciton. In the photocurrent spectrum, the A-2*s* exciton appears at the starting point of a step-like feature and it may result from A-2*s* exciton state merging into the onset of the interband continuum. Figure 2b depicts the photocurrent spectra in log scale as a function of the applied source-drain bias from -1V to -60V that can be converted to the electric field from -0.3V/$\mu m$ to -18.6V/$\mu m$ in the following paragraph. Both A-1*s* and B-1*s* excitons energetically redshift with the increasing in-plane electric field as a result of excitonic DC Stark effect and are well fitted in a quadratic relationship $\sim \Delta E = -\frac{1}{2}\alpha F^2$, where $\alpha$ is the in-plane polarizability, that of A-1*s* is estimated to be around $2 \times 10^{-17} em^2 / V$[17]. Therefore, we could extract the magnitude of in-plane electric field, shown in upper green axis of Figure 2c. Besides, we extract the in-plane polarizability of B-1*s* exciton of 0.75$\alpha_A$ and it concludes that the exciton binding energy of B-1*s* exciton is 1.33 times that of A-1*s* exciton[21], which is consistent with the aforementioned energy difference of A-2*s*/B-2*s* (Fig.2a) and the previous report about a larger effective mass[3][22]. The photocurrent signal below A-1*s* exciton (labeled as trion) totally disappears when V$_{ds}$ is beyond -10V (equivalent to F=-3.1 V/$\mu m$) since the trions cannot survive under a large electric field, as demonstrated in the electric field dependent PL spectra (Fig. 1g).

We also observe two excited exciton states in the vicinity of A-2*s* state, which could not be identified directly. However, we could exclude the possibility of A-2*s* trion state for the peak below A-2*s* for three reasons. Firstly, the aforementioned PL measurement shows that the source-drain voltage generates the in-plane electric field with negligible doping effect. Secondly, throughout the range of electric field where the PL of A-2*s* trion can be probed, the energy of A-2*s* trion has an unresolved shift whereas this peak in the photocurrent spectra redshifts a few meV (Supplementary Note 4). Thirdly, the trion state of A-1$s$ is negligible in the photocurrent experiments under a sizable in-plane electric field. So the trion state of much weaker oscillator strength A-2$s$ is unlikely to be prominent in photocurrent spectra. Figure 2d shows the calculated electro-absorption spectra can mostly reproduce all features including two unidentified peaks from experimental photocurrent spectra. According to the detailed analysis of calculated electro-absorption spectra, we assign them to $2p_x$ ($x$ denotes the direction of in-plane electric field) and 3$s$ states, which is elaborated in the following section.

From the classical dipole model of neutral exciton, there exists a critical field by simply evaluating $F_c = \frac{E_b}{e \cdot r}$ ($E_b, e, r$ are exciton binding energy, electron charge and Bohr radius, respectively) for fully dissociation of Rydberg excitons in monolayer WSe$_2$ (Supplementary Note 5). The actual in-plane electric field with the order of 5V/$\mu m$ is perturbative to 1$s$ exciton while it may be strong enough to ionize excitons with higher principal quantum number ($n \geq 3$). Meanwhile, the electric field of 5V/$\mu m$ is comparable to the critical field for the excited excitons ($n=2$) that include nondegenerate states consisting of 2$s$, 2$p\pm$ excitons with valley degeneracy of two in monolayer TMDs. Therefore, these excitonic states will be influenced enormously based on these assumptions.

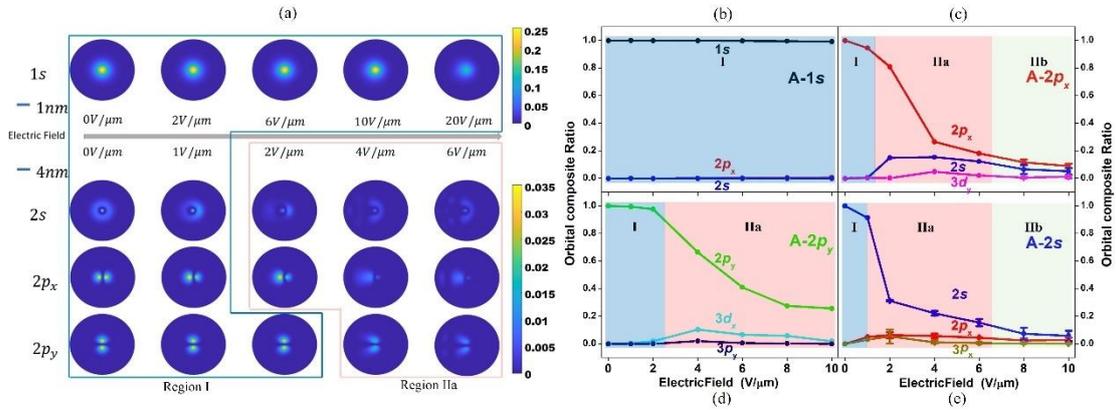

Figure-3 (a) The real spatial distribution of excitonic envelope wavefunction with different orbitals/envelop functions as a function of the in-plane electric field. In region I marked in blue, the exciton states preserve almost unchanged. 1$s$ state keeps unchanged until F=20V/$\mu m$. In region IIa marked in pink, $2p_x, 2p_y$ & 2$s$ states distort evidently owing to the orbital hybridization with some neighbouring states. (b-e) The analysis of the orbital composition of hybrid excitons as a function of in-plane electric field. The error bar denotes the state energy broadening.

To accurately model the exciton energy under an external electric field, we numerically solve the Mott-Wannier exciton model to obtain the eigen-energies and eigen-wavefunctions of Rydberg excitons, the effective Hamiltonian reads:

$$\left\{-\frac{\hbar^2}{2\mu}\nabla^2 + w(r) + e\vec{F}\cdot\vec{r}\right\}\Psi = E\Psi$$

Where the $w(r) = -\frac{e^2}{8\varepsilon_0 r_0}\left[H_0\left(\frac{\kappa r}{r_0}\right) - Y_0\left(\frac{\kappa r}{r_0}\right)\right]$ is the popular Rytova-Keldysh potential describing the screening Coulomb attraction in 2D case, $\mu$, $e\vec{F}\cdot\vec{r}$ represent exciton effective mass and the external electric field term, respectively. The absorption spectra are then calculated as $\omega\, Im(\chi(\omega))$. (See more details in Supplementary Note 6)

We trace the two unidentified peaks under the different in-plane electric fields mentioned in the last section. By calculating the projection of these corresponding eigenfunctions $|\phi\rangle$ on the excitonic states (to distinguish excitonic states whether they are under an in-plane electric field, we define the original excitonic states without electric field by the superscript '0', such as $|1s^0\rangle, |2p_x^0\rangle, |2s^0\rangle$...), we find that the two peaks mainly consist of the $2p_x$ and $3s$ states, respectively, and then we can safely assign them to $2p_x$ and $3s$ excitonic states (more details in Supplementary Note 7). Fig.3a shows the real space distributions of $1s$, $2p_x$, $2p_y$, $2s$-states at various representative electric fields, and Fig.3b-e summarize the dominated orbital compositions of these states. According to the electric field effect on excitons, we describe the scenario exclusively in three regions I, II, III for low, medium, and high electric field, respectively. In the case of low electric field, $e\vec{F}\cdot\vec{r} \ll w(r)$, the external field is perturbative on the excitonic states so that these states almost preserve as original ones. The spatial distributions of these four envelope wavefunctions almost keep unchanged, as demonstrated in region I (the blue frame of Fig. 3a) and the original states keep a major component (more than 90%) in the blue region I of Fig 3b-e. In the region II of medium electric field, we can further divide it into region IIa $e\vec{F}\cdot\vec{r} < w(r)$ and region IIb $e\vec{F}\cdot\vec{r} > w(r)$. In region IIa, the eigenstates are still discrete excitonic states and the spatial distributions of exciton wavefunctions show a finite distortion from the original ones (the pink frame of Fig. 3a). This is because that the excitonic states start to hybridize with some adjacent ones to non-negligible extent, forming hybrid excitons. $2s/2p_x$ state still keeps dominant component in despite of a large reduction in the hybrid exciton shown as the pink region IIa of Fig. 3b-e. $2p_y$ state is relatively immune to the orthogonal electric field ($x$-direction). Here, the dark $2p_x$ state transits from dark to bright state and emerges in the linear photocurrent spectrum as it effectively hybridizes with the bright $2s$ state (Fig.3c). The bright $2s$ state hybridizes with the dark $2p_x$ state at the increase of electric field (Fig. 3e). Note that the dark $2p_y$ state does not hybridize with the $2s$ state, but mixes with the dark $3d_x$ state (Fig.3d). Therefore, it keeps dark and undetectable. In region IIb, $e\vec{F}\cdot\vec{r} > w(r)$, the eigenstates carry similar features as the continuous functions from 2D Airy equation

$\left(-\frac{\hbar^2}{2\mu}\nabla^2 + e\vec{F}\cdot\vec{r}\right)\Psi = E\Psi$). Namely the hybrid excitons mix with the continuous states, leading to effective inhomogeneous broadening so that they become a broad rather than discrete distribution in energy spectrum, denoted as error bars in Fig. 3b-e. Here, we confirm the linewidths of 2$s$ or 2$p_x$ become broader from F=6V/$\mu m$ in the photocurrent experiment (Fig. 2b). Although the 2$s$ or 2$p_x$ component in the hybrid excitons decrease but still give the biggest contribution. In the region III, $e\vec{F}\cdot\vec{r} \gg w(r)$, the screening Coulomb attraction can be considered as a perturbative term of the two-dimensional Airy equation where the eigenstates are mostly continuous. We do not reach so large electric field of the region III in our experiments.

In summary, three-fold in-plane rotational symmetry could theoretically mix the $s$-$p$ excitonic states, but the mixing strength is expected to be too small to brighten 2$p$ state[8,12]. In our case, the applied in-plane electric field is orders of magnitude smaller than the crystal field, but large enough to break the in-plane rotational symmetry of $n$=2 exciton envelope function. Consequently, the angular momentum is not well defined and the wavefunctions of 2$s$ and 2$p$ states are hybridized if the electric field is comparable to the exciton critical field. The dipole-forbidden (dark) A-2$p_x$ excitonic state could be brightened by an in-plane electric field in the linear spectra due to the orbital hybridization with 2$s$ state. It is further confirmed by the simulated absorption spectrum calibrated with the experimental data, as shown in Figure 2d.

The 3$s$ state is predicted to undergo an orbital hybridization with the dark 3$p_x$ & 3$d_y$ states under a small electric field. However, we could not observe 3$s$ state under the electric field ranging from zero until F=-3.1V/$\mu m$ presumably owing to its weak oscillator strength and the signal/noise ratio of the experiment setup. The 3$s$ state clearly emerges under the electric field from -6.2V/$\mu m$ to -15.5 V/$\mu m$ in the photocurrent spectra (Fig.2b). We attribute the enhancement of 3$s$ state to the hybridization with the continuous states. (Supplementary Note 8).

The increase of exciton orbital hybridization is followed by the energy shifts of the hybrid excitons. We record the energy values of hybrid excitons A-1$s$, 2$s$, 2$p_x$, 3$s$ and B-1$s$, 2$s$ from both photocurrent experiments (dots) and the numerical calculation (solid lines, $\mu = 0.23m_e, \kappa = 4.5, r_0 = 4.62nm$ Supplementary Note 5) shown in Figure 4a. They are consistent very well. The 1$s$ exciton experiences a small quadratic redshift with the electric field owing to the negligible orbital hybridization of 2$p_x$ state at F=-18.6V/$\mu m$ (Fig.2c). In contrast, the 2$p_x$ red-shifts while 2$s$ and 3$s$ blue-shift by tens of meV when in-plane electric field goes to 16V/$\mu m$. The 2$p_x$ exciton effectively hybridizes with 2$s$ state while little with 3$d_y$ state. Although it is difficult to resolve A-3$s$ exciton under small electric fields in our experiment, its energy shift under electric fields F=-6.2~ -15.5 V/$\mu m$ is well consistent with the numerical values. In particular, the energies of A-2$s$ and 3$s$ excitons under a fairly strong in-plane electric field could exceed that of the continuous band (1.90eV), denoted as gray dashed line in Fig.4a. This does not mean the exciton binding energy is positive. As the hybrid exciton gains extra electric potential energy from the external electric field, the exciton binding energy is still negative, which secures these exciton states in the photocurrent spectrum. (see Supplementary Note 9)

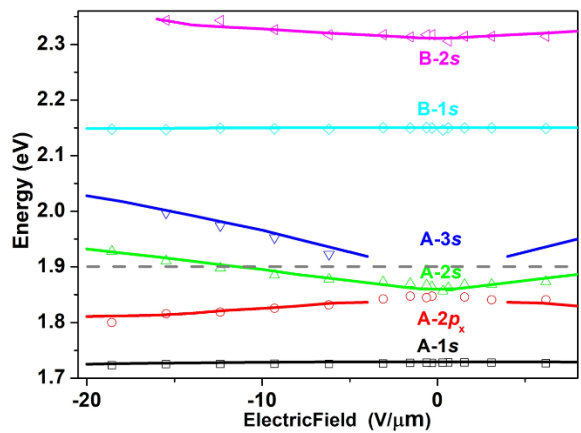

Figure-4 Energies of hybrid excitons from experimental photocurrent spectra and calculated electro-absorption, denoted as dots and solid lines respectively.

In conclusion, the in-plane electric field breaks the in-plane rotational symmetry of the exciton envelope function and mixes exciton states with different excitonic orbitals, resulting in the formation of new hybrid excitons. We demonstrate that the energy, linewidth and oscillator strength of the hybrid excitons can be electrically tuned by the in-plane electric field with the magnitude of 1-16V/$\mu m$. The $2s$ and $2p_x$ excitons can hybridize each other efficiently and this hybridization optically actives the dark $2p_x$ exciton in the one-photon photocurrent spectrum. Besides, the oscillator strength of $3s$ exciton is promoted due its hybridization with continuum band. Also, we observe that the in-plane polarizability of B-$1s$ exciton is smaller than that of A-$1s$ and the energy separation between $2s$ and $1s$ of B exciton is larger than that of A exciton, both implying the higher binding energy of B exciton. Besides, the hybrid excitons under an in-plane electric field inherit the properties of all base orbitals, and consequently they may be bright in both one-photon and two-photon optical transitions. Our results demonstrate the exceptional electric tunability of 2D excitons, providing a potential application of electro-optical modulations with 2D TMDs.

The work was supported by National Natural Science Foundation of China (12104399) and the Hong Kong University Grants Council/ Research grants council under schemes of (AoE/P-701/20), GRF (17300520) and CRF(C7036-17W), AoE seed fund of University of Hong Kong and National Key R&D Program of China (2020YFA0309600). K.W. and T.T. acknowledge support from the Elemental Strategy Initiative conducted by the MEXT, Japan (Grant Number JPMXP0112101001) and JSPS KAKENHI (Grant Numbers 19H05790 and JP20H00354).